\documentclass[12pt, a4paper, twoside]{article}
\usepackage{graphicx}
\usepackage{color}
\usepackage{lscape}
\usepackage{bm}
\newcommand*{\B}[1]{\ifmmode\bm{#1}\else\textbf{#1}\fi}
\usepackage{epstopdf}
\usepackage{lscape}
\newenvironment{changemargin}[2]{%
\begin{list}{}{%
\setlength{\leftmargin}{#1}%
\setlength{\rightmargin}{#2}%
}%
\item[]}
{\end{list}}
\usepackage{graphicx}
\usepackage{color}

\usepackage{epstopdf}
\usepackage{pdflscape}
\usepackage{amsmath}
\usepackage{amsmath}
\usepackage{amsmath}
\usepackage{kantlipsum}
\allowdisplaybreaks
\usepackage{setspace}

\usepackage{array}

\newcolumntype{M}[1]{>{\raggedright\arraybackslash}m{#1}}
\newcolumntype{N}{@{}m{0pt}@{}}
\usepackage{amsmath}
\usepackage{setspace}
\begin{document}
\baselineskip=0.27in
{\bf \LARGE
\begin{changemargin}{-0.5cm}{-0.5cm}
\begin{center}
{An electron of helium atom under a high-intensity laser field}\footnote{\bf To appear in Laser Phys. 2017.}
\end{center}
\end{changemargin}}
\vspace{4mm}
\begin{center}
\large{\bf Babatunde James Falaye$^{a, b, }$}\footnote{\scriptsize E-mail:~ fbjames11@physicist.net;~ babatunde.falaye@fulafia.edu.ng}, {\large{\bf Guo-Hua Sun$^{c, }$}\footnote{\scriptsize E-mail:~ sunghdb@yahoo.com}}, \large{\bf Adenike Grace Adepoju$^{d,}$}\footnote{\scriptsize E-mail:~ aagrace06@gmail.com}, {\large{\bf Muhammed S. Liman$^{b, }$}}\footnote{\scriptsize E-mail:~ limanms2@gmail.com}, {\large{\bf K. J. Oyewumi$^{e, }$}}\footnote{\scriptsize E-mail:~ kjoyewumi66@unilorin.edu.ng},  \large{\bf and} {\large{\bf Shi-Hai Dong$^{d,}$}}\footnote{\scriptsize E-mail:~ dongsh2@yahoo.com}
\end{center}
{\footnotesize
\begin{center}
{\it $^\textbf{a}$Departamento de F\'isica, Escuela Superior de F\'isica y Matem\'aticas, Instituto Polit\'ecnico Nacional, Edificio 9, Unidad Profesional Adolfo L\'opez Mateos, Mexico D.F. 07738, Mexico.} {\it $^\textbf{b}$Applied Theoretical Physics Division, Department of Physics, Federal University Lafia,  P. M. B. 146, Lafia, Nigeria.} {\it $^\textbf{c}$Catedr\'atica CONACyT, CIC, Instituto Polit\'{e}cnico Nacional, Unidad Profesional Adolfo L\'opez Mateos, Mexico D. F. 07700, Mexico.}{\it $^\textbf{d}$CIDETEC, Instituto Polit\'{e}cnico Nacional, Unidad Profesional Adolfo L\'opez Mateos, M\'{e}xico D. F. 07700, M\'{e}xico.} {\it $^\textbf{e}$Theoretical Physics Section, Department of Physics, University of Ilorin,  P. M. B. 1515, Ilorin, Nigeria.} 
\end{center}}

\begin{abstract}
\noindent
We scrutinize the behavior of eigenvalues of an electron of Helium atom as it interacts with electric field directed along $z$-axis and exposed to linearly polarized intense laser field radiation. In order to achieve this, we freeze one electron of the helium atom at its ionic ground state and the motion of the second electron in the ion core is treated via a more general case of screened Coulomb potential model. Using the Kramers-Henneberger (KH) unitary transformation, which is semiclassical counterpart of the Block-Nordsieck transformation in the quantized field formalism, the squared vector potential that appears in the equation of motion is eliminated and the resultant equation is expressed in KH frame. Within this frame, the resulting potential and the corresponding wave function have been expanded in Fourier series and using Ehlotzkys approximation, we obtain a laser-dressed potential to simulate intense laser field. By fitting the more general case of screened Coulomb potential model into the laser-dressed potential, and then expanding it in Taylor series up to $\mathcal{O}(r^4,\alpha_0^9)$, we obtain the solution (eigenvalues and wave function) of an electron of Helium atom under the influence of external electric field and high-intensity laser field, within the framework of perturbation theory formalism. We found that the variation in frequency of laser radiation has no effect on the eigenvalues of an electron of helium for a particular electric field intensity directed along $z$-axis. Also, for a very strong external electric field and an infinitesimal screening parameter, the system is strongly bound. This work has potential application in the areas of atomic and molecular processes in external fields including interactions with strong fields and short pulses.
\end{abstract}
\noindent
{\bf PACs Nos}: 34.50.Fa, 52.25.Kn, 52.27.Gr, 52.27.Lw

\section{Introduction}
Lasers have emerged as one of the world's indispensable technologies, employed in telecommunications, law enforcement, military equipments, etc. Laser pulses control various atomic or molecular process. For instance, atoms undergo about three ionization when probed by a laser of controlled intensity. Recent furtherance in laser technology has aroused the interest of many researchers to investigate new sources of laser in order to probe and control molecular structure, function and dynamics on the natural timescale of atomic motion, the femtosecond and electron motion on attosecond timescale \cite{EF1}. To obtain an intense laser field, it is required to concentrate large amounts of energy within short period of time, and then focus the laser light onto a small area. In an intense laser system, a train of pulses of short duration are created by the oscillator. The energy of the pulses is then proliferated by the amplifier, which are eventually focused.

Studying atoms in intense laser field have been a subject of active research for more than three decades due to its salient application in the invention of high-power short-pulse laser technologies. These atoms exhibit new properties that have been discovered via the study of multiphoton processes \cite{EF2}. When a high-power laser is directed into a gas of atoms, the magnitude of electromagnetic field is found to be consistent with the Coulomb field, which binds a 1s electron in a Hydrogen atom \cite{EF3}. Within this context, so many outstanding results focusing on hydrogen atom have been reported so far (see \cite{EF4,EF5,EF6,EF7,EF8} and refs. therein). It was shown in ref.  \cite{EF9}, that in the presence of an oscillating magnetic field, the ionization rate of a hydrogen atom interacting with intense laser  dwindle, and the electron density becomes ionized with a less rate by keeping the magnetic field strength constant and increasing the intensity of the laser.

On the other hand, studying helium atom under intense laser field has also received great attention from both theorists and experimentalists \cite{EF10,EF11,EF12,EF13}. Chattaraj \cite{EF14} studied dynamic response of a helium atom in an intense laser field within a framework of quantum fluid density. Chen et al. \cite{EF15} numerically simulated the double-to-single ionization ratio for the helium atom under intense laser field. In fact electron-helium scattering in the presence of laser field was recently reported in ref. \cite{EF16}. So many outstanding contributions have been made on this subject, however, it is worth mentioning that to our best knowledge, there has been no account of the current study both experimentally and theoretically. In the present work, our main focus is to scrutinize the behavior of eigenvalues of an electron of Helium atom under the influence of external electric field and exposed to linearly polarized intense laser field radiation. This study will be of great interest in the areas of atomic and molecular processes in external fields including interactions with strong fields and short pulses.

\section{Formulation of the problem}
In this section, we derive the equation of motion for spherically confined one electron of Helium atom under the influence of external electric field directed along $z$-axis and exposed to linearly polarized intense laser field radiation. Achieving our goal in this section requires that we start with the following time-dependent Schr\"odinger wave equation
\begin{equation}
i\hbar\frac{\partial}{\partial t}\Psi(\bm{r},t)=\left[-\frac{\hbar^2}{2\mu}{\nabla}^2-i\hbar\frac{e}{2\mu}\big[\bm{A}(\bm{r},t)\cdot{\nabla}+{\nabla}\cdot {\bf A}(\bm{r},t)\big]+\frac{e^2}{2\mu}{\bf A}(\bm{r},t)^2-e\phi+V(\bm{r})-Fr\right]\Psi(\bm{r},t),
\label{EQ1}
\end{equation}
with the scalar potential $\phi({\bf r},t)$ and the vector potential $\bm{A}(\bm{r},t)$ which is invariant under the gauge transformation. $\mu$ is the effective mass of the electron. Furthermore, $Fr$ describes a radial electric field.  We consider Coulomb gauge, such that $\nabla\cdot {\bf A}(\bm{r},t)=0$  with $\phi=0$ in empty space and then simplify the interaction term in the equation (\ref{EQ1}) by performing gauge transformations within the framework of dipole approximation. In this approximation, for an atom whose nucleus is located at the position $r_0$, the vector potential is spatially homogeneous ${\bf A}(\bm{r},t)\approx\bm{A}(t)$. Moreover, term ${\bf A}({\bf r},t)^2$ appearing in equation (\ref{EQ1}) is considered for extremely high field strength. It is usually small and can be eliminated by extracting a time-dependent phase factor from the wave function via \cite{EF17}
\begin{equation}
\Psi^{v}(\bm{r},t)=\exp\left[\frac{ie^2}{2\mu\hbar}\int_{-\infty}^t{\bf A}(t')^2dt'\right]\Psi(\bm{r},t),
\label{EQ2}
\end{equation}
to obtain {velocity gauge}\footnote{Because the vector potential $A(t)$ is being coupled to the operator $\bm{p}/m$ via the interaction Hamiltonian, where $\bm{p}=-i\hbar{\nabla}.$}
\begin{equation}
i\hbar\frac{\partial}{\partial t}\Psi^{v}({\bf r},t)=\left[-\frac{\hbar^2}{2\mu}\nabla^2-i\hbar\frac{e}{\mu}{\bf A}(t)\cdot\nabla+V(r)-Fr\right]\Psi^{v}({\bf r},t).
\label{EQ3}
\end{equation}
A prerequisite to study an electron of helium atom under intense high-frequency laser field is transforming equation (\ref{EQ3}) to the Kramers-Henneberger accelerated frame. Now, with the introduction of the following unitary Kramers-Henneberger's transformation
\begin{equation}
\Psi^{A}({\bf r},t)= U^\dag\Psi^{v}({\bf r},t)\ \mbox{with} \ \ \ U=\exp\left[-\frac{i}{\hbar}\bm{\alpha}(t).\bm{p}\right],\ \ \mbox{and} \ \bm{\alpha}(t)=\frac{e}{\mu}\int^t{\bf A}(t')dt',
\label{EQ4}
\end{equation}
which is semiclassical counterpart of the Block-Nordsieck transformation in the quantized field formalism, the coupling term $\bm{A}(t)\cdot\bm{p}$ in the velocity gauge (i.e., Eq. (\ref{EQ3})) is eliminated. More explicitly, this can be done via
\begin{equation}
i\hbar U^\dag\frac{\partial}{\partial t}U\Psi^{A}(\bm{r},t)=U^\dag\left[-\frac{\hbar^2}{2\mu}\nabla^2-i\hbar\frac{e}{\mu}{\bf A}(t)\cdot{\nabla}+V(\bm{r})-Fr\right]U\Psi^{A}({\bf r},t).
\label{EQ5}
\end{equation}
Evaluation of terms in equation (\ref{EQ5}) are straightforward and easy. However, let us try to be more explicit in evaluating the term $U^\dag V(r)U$. This can be done via Campbell-Baker-Hausdorff identity: $e^{\hat{A}}\hat{B}e^{-\hat{A}}=\hat{B}+[\hat{A},\hat{B}]+[\hat{A},[\hat{A},\hat{B}]]/2!+\ldots$. Thus, we have
\begin{eqnarray}
U^\dag V(r)U&=&\exp\left[\frac{i}{\hbar}\bm{\alpha}(t).\bm{p}\right]V(\bm{r})\exp\left[-\frac{i}{\hbar}\bm{\alpha}(t).\bm{p}\right]\nonumber\\
&=&V(\bm{r})+\left[\bm{\alpha}(t).\nabla\right]V(\bm{r})+\frac{1}{2!}\left[\bm{\alpha}(t).\nabla\right]^2V(\bm{r})+\ldots\nonumber\\
&=&V\left[r+\bm{\alpha}(t)\right],
\label{EQ6}
\end{eqnarray}
where $\bm{\alpha}(t)$ denotes the displacement of a free electron in the incident laser field. Hence, Eq. (\ref{EQ5}) becomes
\begin{equation}
i\hbar\frac{\partial}{\partial t}\Psi^{A}(\bm{r},t)= \left[-\frac{\hbar^2}{2\mu}\nabla^2+V\left[\bm{r}+\bm{\alpha}(t)\right]-Fr\right]\Psi^{A}(\bm{r},t).
\label{EQ7}
\end{equation}
Eq. (\ref{EQ7}) represents a space-translated version of the time-dependent Schr\"odinger wave equation with incorporation of $\bm{\alpha}(t)$ into the potential in order to simulate the interaction of atomic system with the laser field. It is worth mentioning that K-H transformation leaves the term $-Fr$ invariant. Now, for a steady field condition, the vector potential takes the form $\bm{A}(t)=(\mathcal{E}_0/\omega)\cos(\omega t)$ with $\bm{\alpha}(t)=\alpha_0\sin{(\omega t)}$, where $\alpha_0=e\mathcal{E}_0/(\mu\omega^2)$ is the amplitude of oscillation of a free electron in the field (called as {\it laser-dressing parameter}), $\mathcal{E}_0$ denotes the amplitude of electromagnetic field strength and $\omega$ is the angular frequency. Now, considering a pulse where the electric field amplitude is steady, the wave function in the frame of Kramers-Henneberger takes the following Floquet form \cite{EF17}:
\begin{equation}
\Psi^{A}(\bm{r},t)=e^{-\frac{iE_{_{KH}}}{\hbar}t}\sum_{n}\Psi_n^{E_{_{KH}}}(\bm{r})e^{-in\omega t},
\label{EQ8}
\end{equation}
where Floquet quasi-energy has been denoted by $E_{KH}$. The potential in the frame of Kramers-Henneberger can be expanded in Fourier series as \cite{EF18}
\begin{equation}
V[\bm{r}+\bm{\alpha}(t)]=\sum_{m=-\infty}^\infty V_m(\alpha_0;\bm{r})e^{-im\omega t}\ \ \ \mbox{with}\ \ \ V_m(\alpha_0;\bm{r})=\frac{i^m}{\pi}\int_{-1}^{1} V(\bm{r}+{\alpha}_0\varrho)\frac{T_n(\varrho)}{\sqrt{1-\varrho^2}}d\varrho,
\label{EQ9}
\end{equation}
where we have taken the period as $2\pi/\omega$ and introduced a new transformation of the form $\varrho=\sin(\omega t)$. Furthermore, $T_n(\varrho)$ are Chebyshev polynomials. Substituting Eqs. (\ref{EQ8}) and (\ref{EQ9}) into Eq. (\ref{EQ7}), yields a set of coupled differential equation:
\begin{equation}
\left[-\frac{\hbar^2}{2\mu}\nabla^2+V_m(\alpha_0;\bm{r})-Fr-(E_{KH}+n\hbar\omega)\right]\Psi_n^{E_{_{KH}}}(\bm{r})=-\sum_{m=-\infty}^{\infty, m\neq n} V_{n-m}\Psi_m^{E_{_{KH}}}(\bm{r}).
\label{EQ10}
\end{equation}
Considering $n=0$ (which gives the lowest order approximation) and high frequency limit (which made $V_{m}$ with $m\neq0$ vanish), Eq. (\ref{EQ10}) becomes
\begin{equation}
\left[-\frac{\hbar^2}{2\mu}\nabla^2+V_0(\alpha_0;\bm{r})-Fr-E_{KH}\right]\Psi_0^{E_{_{KH}}}(\bm{r})=0.
\label{EQ11}
\end{equation}
and the coefficient of the Fourier series for the potential becomes
\begin{eqnarray}
V_0(\alpha_0;\bm{r})&=&\frac{1}{\pi}\int_{-1}^{1}V(\bm{r}+\alpha_0\varrho)\frac{d\varrho}{\sqrt{1-\varrho^2}}
=\frac{1}{\pi}\int_{0}^{1}\left[V(\bm{r}+\alpha_0\varrho)+V(\bm{r}-\alpha_0\varrho)\right]\frac{d\varrho}{\sqrt{1-\varrho^2}}.
\label{EQ12}
\end{eqnarray}
Using Ehlotzky approximation \cite{EF19}, one has $[V(\bm{r}+\alpha_0\varrho)+V(\bm{r}-\alpha_0\varrho)]\approx [V(\bm{r}+\alpha_0)+V(\bm{r}-\alpha_0)]$. Hence, by evaluating the integral, we obtain
\begin{eqnarray}
V_0(\alpha_0;\bm{r})=\frac{1}{2}\left[V(\bm{r}+\alpha_0)+V(\bm{r}-\alpha_0)\right].
\label{EQ13}
\end{eqnarray}
Eq. (\ref{EQ13}) is the approximate expression to model laser field. Now, let us incorporate the model potential to simulate the behavior of one electron of helium atom into model potential (\ref{EQ13}). In order to model this potential, we freeze one electron of the helium at its ground state and then consider the motion of the second electron in the ion core. We would like to advice the readers to check literature \cite{EF11,EF20} for more details about this. Moreover, an appropriate model potential for this has been presented and studied in ref. \cite{EF20}: $V(r)=-(a/r)[1+(1+br)e^{-2br}]$. Hence, Eq. (\ref{EQ11}) becomes
\begin{equation}
\left[-\frac{\hbar^2}{2\mu}\nabla^2-\frac{a}{r_{\alpha_0^{+}}}\left[1+(1+br_{\alpha_0^{+}})e^{-2br_{\alpha_0^{+}}}\right]-\frac{a}{r_{\alpha_0^{-}}}\left[1+(1+br_{\alpha_0^{-}})e^{-2br_{\alpha_0^{-}}}\right]-Fr-E_{KH}\right]\Psi_0^{E_{_{KH}}}(\bm{r})=0.
\label{EQ14}
\end{equation}
where $r_{\alpha_0^{\pm}}=r\pm\alpha_0$, $a$ denotes the strength coupling constant and $b$ represents the screening parameter.  The above equation (\ref{EQ14}) is the equation of motion for spherically confined one electron of helium atom exposed to linearly polarized intense laser field radiation. Now, in order to achieve the objective of this study, in the next section, we solve Eq. (\ref{EQ14}) within the framework of perturbation formalism.

\begin{figure*}[!t]
\centering \includegraphics[height=140mm, width=165mm]{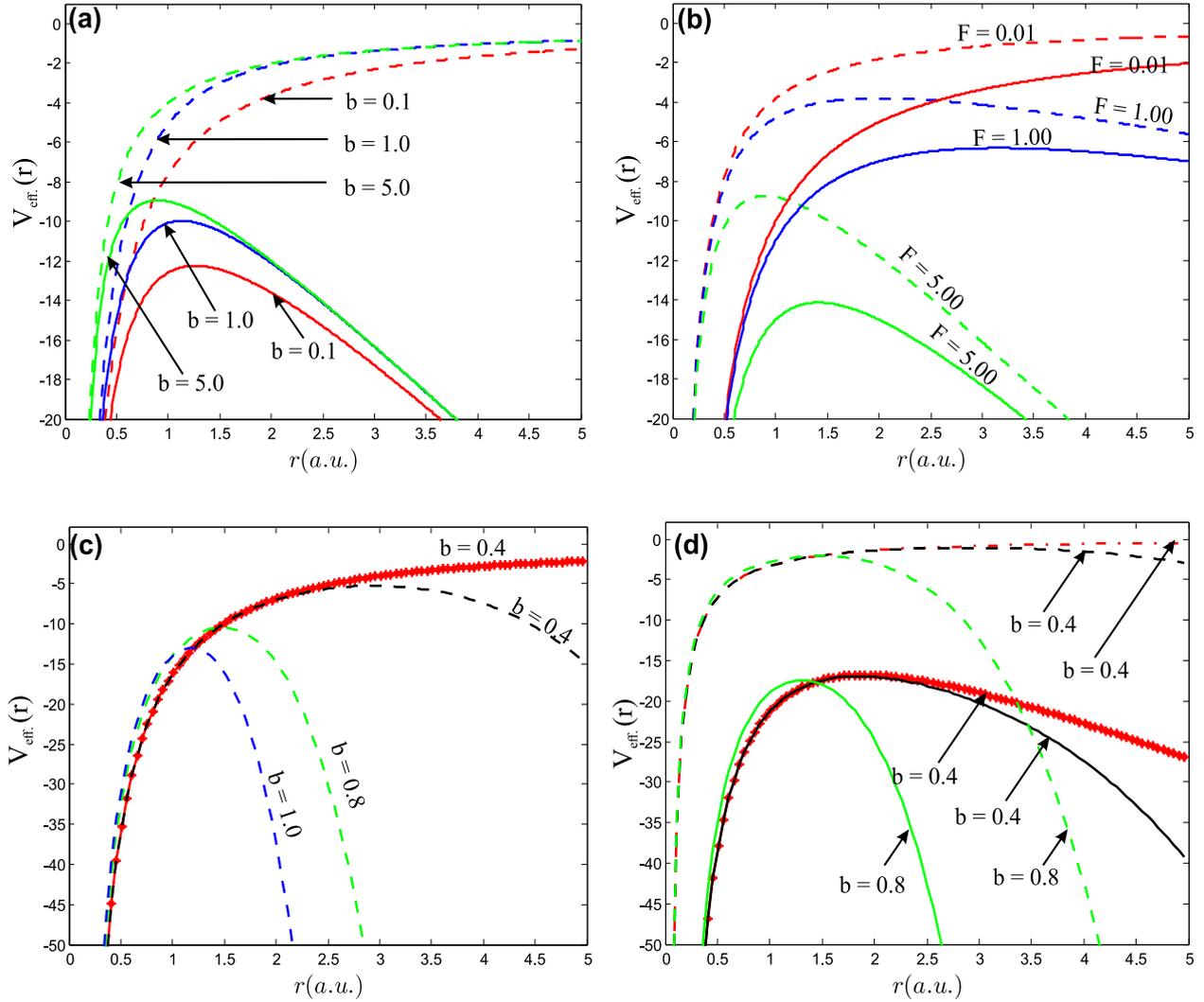}
\caption{\protect\footnotesize The plots of model potential as a function of internuclear distance. In figure (a), we depict the shape of the model potential by considering a weak and strong external electric field via taking $F$ as $0.01$ and $5$ respectively for various values of screening parameter $b$ with a fixed coupling strength $a=1$. The dash line ``- -" represents the context of weak $F$ while the thick line represents strong $F$. Furthermore, in figure (b),  we depict the shape of the model potential by taking screening parameter \& coupling strength as 1 \& 0.1 (dash lines) and 5 \& 5 (thick lines) respectively for various values of electric field strengths. In figure (c), taking $a=5$, we show the accuracy of the approximate expression for the potential model (i.e., the Taylor series expansion of the effective potential) given in Eq. (\ref{EQ16}) for various values of screening parameter under a weak external electric field $F=0.01$. The line with  marker type ``$-*$" represents the effective potential while the dot-dash lines ``$\cdot -$" denote the approximate expression for various values of screening parameter. Figure (d) explains figure (c) further, to see if coupling strength and external electric field affects the approximate expression. The lines with  marker type ``$-*$" and ``$\cdot -$" represents the effective potential for $ (a, F)=(5, 5)$ and  $(a, F)=(1, 0.01)$ respectively while the lines marker ``$-$" and ``$--$", denote the approximate expression for various values of screening parameter within the frame of $(a, F)=(5, 5)$ and  $(a, F)=(1, 0.01)$ respectively. We have taken $\alpha_0=0.001$ and all our numerical computations are in atomic units (a.u.). It's also important to note that we have used MATLAB for our computations and the line marker notations are in respect to this software.}
\label{fig1}
\end{figure*}
\section{Eigenspectra Calculation}
Eq. (\ref{EQ14}) is not solvable analytically. One can either use numerical procedure or perturbation formalism. Using perturbation approach, we decompose the equation into two parts where the first part is exactly solvable and the other part is perturbation. Consequently, the eigenvalue solutions are represented in power series with the leading term corresponding to the solution of exactly solvable part and the other part is correction to the energy term which corresponds to the perturbation term. This approach has been used in numerous research reports (See \cite{EF21,EF22} and references therein). Now, we re-write Eq. (\ref{EQ14}) as
\begin{equation}
\frac{\hbar^2}{2\mu}\left(\frac{\nabla^2\mathcal{X}_0(r)}{\mathcal{X}_0(r)}+\frac{\nabla^2\mathcal{Y}_0(r)}{\mathcal{Y}_0(r)}+2\frac{\nabla\mathcal{X}_0(r)\nabla\mathcal{Y}_0(r)}{\mathcal{X}_0(r)\nabla\mathcal{Y}_0(r)}\right)=V_{\rm eff.}(r)-E_{KH},
\label{EQ15}
\end{equation}
where $\Psi_0^{E_{_{KH}}}(\bm{r})=\mathcal{X}_0(r)\mathcal{Y}_0(r)$ with $\mathcal{X}_0(r)$ as the wave function of the exactly solvable part and $\mathcal{Y}_0(r)$ as the moderating wave function. The effective potential $V_{\rm eff.}(r)$ represents the Taylor's series expansion of the potential terms in Eq. (\ref{EQ14}). This can be written as:
\begin{eqnarray}
V_{\rm eff.}(r)&=& -\frac{4a}{r}+\left(2ab-\frac{4}{405}ab^9\alpha_0^8-\frac{8}{63}ab^7\alpha_0^6-\frac{4}{5}ab^5\alpha_0^4-\frac{4}{3}ab^3\alpha_0^2\right)\nonumber\\
&&+\left(\frac{32}{1575}ab^{10} \alpha_0^8+\frac{4}{15}ab^8\alpha_0^6+\frac{16}{9}ab^6\alpha_0^4+4ab^4\alpha_0^2-F\right)r\nonumber\\
&&+\left(-\frac{8}{385}ab^{11}\alpha_0^8-\frac{112}{405}ab^9\alpha_0^6-\frac{40}{21}ab^7\alpha_0^4-\frac{24}{5}a b^5\alpha_0^2-\frac{4}{3}ab^3\right)r^2\nonumber\\
&&+\left(\frac{8}{567}ab^{12}\alpha_0^8+\frac{128}{675}ab^{10}\alpha_0^6+\frac{4}{3}ab^8\alpha_0^4+\frac{32}{9}ab^6\alpha_0^2+\frac{4}{3}ab^4\right)r^3\nonumber\\
&&+\left(-\frac{88}{12285}ab^{13}\alpha_0^8-\frac{16}{165}ab^{11}\alpha_0^6-\frac{56}{81}ab^9\alpha_0^4-\frac{40}{21}a b^7\alpha_0^2-\frac{4ab^5}{5}\right)r^4+\mathcal{O}(r^5,\alpha_0^9).\nonumber\\
\label{EQ16}
\end{eqnarray}
The first term is the main part which corresponds to a shape invariant potential for which the superpotential is known analytically and the remaining part is taken as a perturbation, $\Delta V_{\rm eff.}(r)$. At this junction, one may be intrigued about why the series have been truncated at fourth order of $r$. Apropos of this, it should be noted that convergence is not an important property for series approximations in physical problems. A slowly convergent approximation that requires many terms to achieve reasonable accuracy is much less valuable than the divergent series, which gives accurate answers in a few terms.

\begin{figure*}[!h]
\centering \includegraphics[height=130mm, width=155mm]{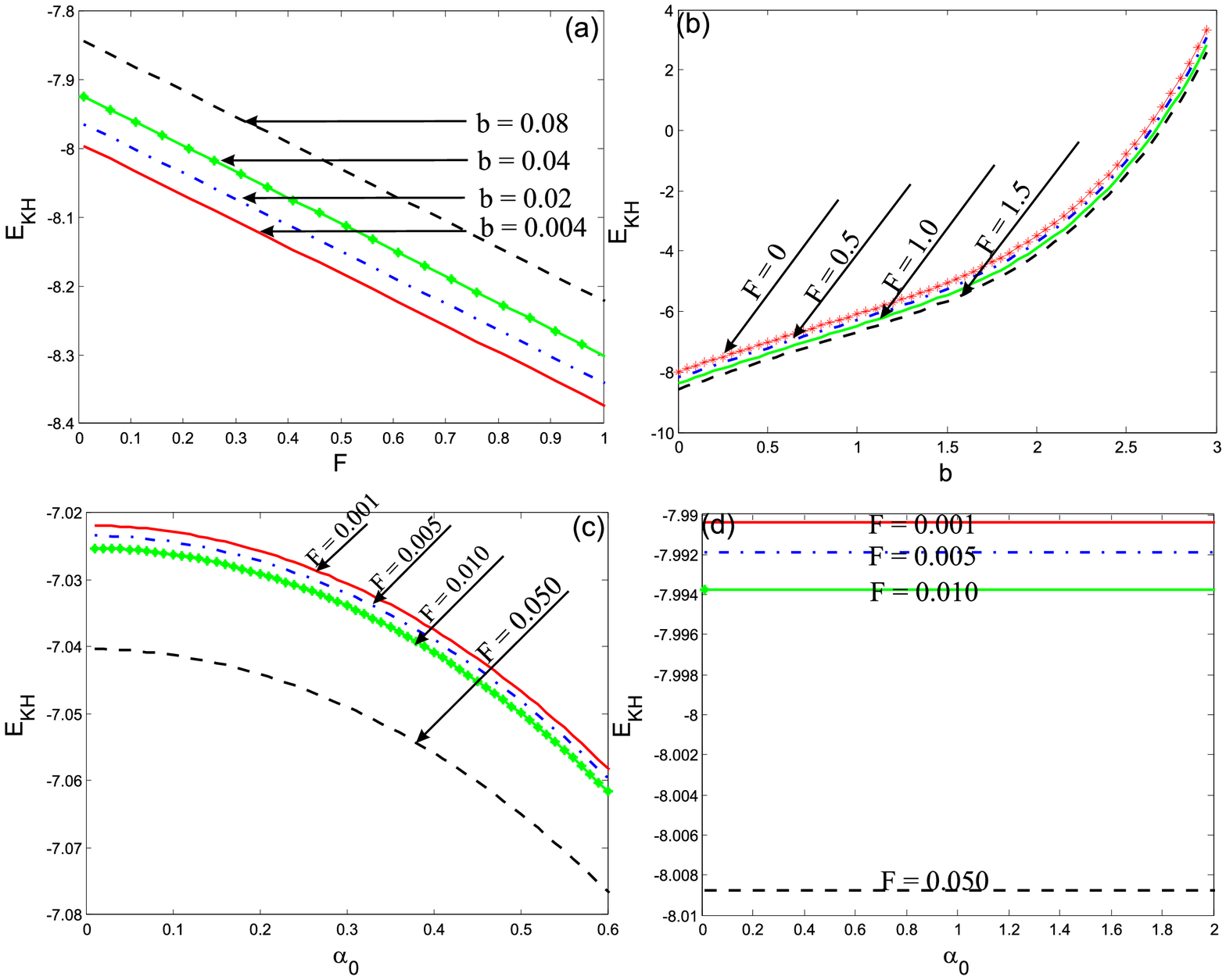}
\caption{\protect\footnotesize Plots of energy eigenvalues of an electron of helium atom interacting with electric field and exposed to linearly polarized intense laser field radiation, as a function of various model parameters. (a) Eigenvalues as a function of electric field with $\alpha_0=0.001$ and for various values of screening parameter. (b) Eigenvalues as a function of screening parameter with $\alpha_0=0.001$ and for various intensities of electric field. (c) Energy levels as a function of laser-dressing parameter with $b=0.5$ for various intensities of electric field. (d) Same as (c) but for $b=0.005$. All our numerical computations are in a.u..}
\label{fig2}
\end{figure*}

Now, taking the logarithmic derivatives of the perturbed and unperturbed wave functions as $W_0(r)=-(\hbar/\sqrt{2\mu})\mathcal{X}_0'/\mathcal{X}_0$ and $\Delta W_0(r)=-(\hbar/\sqrt{2\mu})\mathcal{Y}_0'/\mathcal{Y}_0$, and then substitute them into (\ref{EQ15}), yield the following equation
\begin{subequations}
\begin{equation}
\frac{\hbar^2}{2\mu}\frac{\mathcal{X}_0''(r)}{\mathcal{X}_0(r)}=W_0^2(r)-\frac{\hbar}{\sqrt{2\mu}}W_0'(r)= -\frac{4a}{r}-E_{KH}^{(0)},
\label{EQ17a}
\end{equation}
\begin{equation}
\Delta W_0^2(r)-\frac{\hbar}{\sqrt{2\mu}}\Delta W_0'(r)+2W_0(r)\Delta W_0(r)=\Delta V_{\rm eff.}(r)-\Delta E_{KH},
\label{EQ17b}
\end{equation}
\end{subequations}
where $E_{KH}^{(0)}$ is the eigenvalue of exactly solvable part and $\Delta E_{KH}=E_{KH}^{(1)}+E_{KH}^{(2)}+E_{KH}^{(3)}+...$ is correction to the energy which corresponds to the perturbation term. Eq. (\ref{EQ17a}) is analytically solvable via formula method \cite{EF23} to obtain
\begin{equation}
\mathcal{X}_0(r)=2\zeta^{3/2}re^{-\zeta r},\ \ \ W_0(r)=-\frac{\hbar}{r\sqrt{2\mu}}+\frac{2a\sqrt{2\mu}}{\hbar},\ \ \ E_{KH}^{(0)}=-2a\zeta,\ \ \mbox{where}\ \ \zeta=\frac{4a\mu}{\hbar^2}.
\label{EQ18}
\end{equation}
On the contrary, Eq. (\ref{EQ17b}) is not exactly solvable. It is therefore required to expand the related functions as
$\Delta V_{\rm eff.}(r;\eta)=\sum_{i=1}^\infty\eta_iV_{\rm eff.}(r)^{(i)}$, $\Delta W_{0}(r;\eta)=\sum_{i=1}^\infty\eta_iW_{0}^{(i)}$, $\Delta E_{0}^{(i)}(\eta)=\sum_{i=1}^\infty\eta_iE_{0}^{(i)}$, where $i$ represents the order of perturbation. We substitute these expressions into equation (\ref{EQ17b}) and then equate terms with same power of $\eta$ on both sides to have the following expressions
\begin{subequations}
\begin{eqnarray}
2W_{0}(r)W_{0}^{(1)}(r)-\frac{\hbar}{\sqrt{2\mu}}\frac{dW_{0}^{(1)}(r)}{dr}=V_{\rm eff.}^{(1)}(r)-E_{KH}^{(1)}, \label{EQ19a}\\
\left[W_{0}^{(1)}(r)\right]^2+2W_{0}(r)W_{0}^{(2)}(r)-\frac{\hbar}{\sqrt{2\mu}}\frac{dW_{0}^{(2)}(r)}{dr}=V_{\rm eff.}^{(2)}(r)-E_{KH}^{(2)}, \label{EQ19b}\\
2\left[W_{0}(r)W_{0}^{(3)}(r)+W_{0}^{(1)}(r)W_{0}^{(2)}(r)\right]-\frac{\hbar}{\sqrt{2\mu}}\frac{dW_{0}^{(3)}(r)}{dr}=V_{\rm eff.}^{(3)}(r)-E_{KH}^{(3)}, \label{EQ19c}\\
2\left[W_{0}(r)W_{0}^{(4)}(r)+W_{0}^{(1)}(r)W_{0}^{(3)}(r)\right]+W_{0}^{(2)}(r)W_{0}^{(2)}(r)-\frac{\hbar}{\sqrt{2\mu}}\frac{dW_{0}^{(4)}(r)}{dr}=V_{\rm eff.}^{(4)}(r)-E_{KH}^{(4)}.
\label{EQ19d}
\end{eqnarray}
\end{subequations}
Taking the superpotentials into account and then multiplying each term in equations (\ref{EQ19a}-\ref{EQ19d}) by $\mathcal{X}_0^2(r)$, we obtain first, second and third -order corrections to the energy and their superpotentials as follows:
\begin{subequations}
\begin{eqnarray}
E_{KH}^{(1)}&=&\int_{0}^{\infty}\mathcal{X}_{0}^2(r)r\left(\frac{32}{1575}ab^{10} \alpha_0^8+\frac{4}{15}ab^8\alpha_0^6+\frac{16}{9}ab^6\alpha_0^4+4ab^4\alpha_0^2-F\right)dr\nonumber\\
&=&\frac{\hbar^2b^2}{2\mu}\left(\frac{8}{525}b^{8}\alpha_0^8+\frac{1}{5}b^6\alpha_0^6+\frac{4}{3}b^4\alpha_0^4+3b^2\alpha_0^2-\frac{3F}{4ab^2}\right)
\label{EQ20a}\\
W_{0}^{(1)}(r)&=&\sqrt{\frac{2\mu}{\hbar^2}}\frac{1}{\mathcal{X}_{0}^2(r)}\int_{}^{r}\mathcal{X}_{0}^2(\varrho)\left[E_{KH}^{(1)}-\left(\frac{32}{1575}ab^{10} \alpha_0^8+\frac{4}{15}ab^8\alpha_0^6+\frac{16}{9}ab^6\alpha_0^4+4ab^4\alpha_0^2-F\right)\varrho\right]d\varrho\nonumber\\
&=&\frac{rb^2\hbar}{\sqrt{2\mu}}\left(\frac{8}{1575}b^{8} \alpha_0^8+\frac{1}{15}b^6\alpha_0^6+\frac{4}{9}b^4\alpha_0^4+b^2\alpha_0^2-\frac{F}{4ab^2}\right)\label{EQ20b}\\
E_{KH}^{(2)}&=&\int_{0}^{\infty}\mathcal{X}_{0}^2(r)\bigg[r^2 \left(-\frac{8}{385}ab^{11}\alpha_0^8-\frac{112}{405}ab^9\alpha_0^6-\frac{40}{21}ab^7\alpha_0^4-\frac{24}{5}a b^5\alpha_0^2-\frac{4}{3}ab^3\right)\nonumber\\
&&\ \ \ \ \ \ \ \ \ \ \ \ \ \ \ \ \ \ \ \ \ \ \ \ \ \ \ \ \ \ \ \ \ \ \ \ \ \ \ -\left[{W_{0}^{(1)}}(r)\right]^2\bigg]dr\nonumber\\
&=&\bigg[\frac{\hbar^4b^3}{\mu^2a}\left(-\frac{3}{770}b^{8}\alpha_0^8-\frac{7}{135}b^6\alpha_0^6-\frac{5}{14}4^7\alpha_0^4-\frac{9}{10} b^2\alpha_0^2-\frac{1}{4}\right)\nonumber\\
&&\ \ \ \ \ \ \ \ \ \ \ \ \ \ \ \ \ \ \ \ \ -\frac{3b^2\hbar^6}{32\mu^3a^2}\left(\frac{8}{1575}b^{8} \alpha_0^8+\frac{1}{15}b^6\alpha_0^6+\frac{4}{9}b^4\alpha_0^4+b^4\alpha_0^2-\frac{F}{4ab^2}\right)^2\bigg]\label{EQ20c}\\
W_{0}^{(2)}(r)&=&\sqrt{\frac{2\mu}{\hbar^2}}\frac{1}{\mathcal{X}_{0}^2(r)}\bigg[\int_{}^{r}\mathcal{X}_{0}^2(\varrho)\left(E_{KH}^{(2)}+\left[W_{0}^{(1)}(\varrho)\right]^2\right)d\varrho.\nonumber\\
&&\ \ \ \  -\int_{}^{r}\mathcal{X}_{0}^2(\varrho)\left(-\frac{8}{385}ab^{11}\alpha_0^8-\frac{112}{405}ab^9\alpha_0^6-\frac{40}{21}ab^7\alpha_0^4-\frac{24}{5}a b^5\alpha_0^2-\frac{4}{3}ab^3\right)\varrho^2d\varrho\bigg]\nonumber\\
&=&\frac{rb^3\hbar}{\sqrt{2\mu}}\left(r+\frac{\hbar^2}{2\mu a}\right)\left(-\frac{2}{385}b^{8}\alpha_0^8-\frac{28}{405}b^6\alpha_0^6-\frac{10}{21}b^4\alpha_0^4-\frac{6}{5} b^2\alpha_0^2-\frac{1}{3}\right)\nonumber\\
&&\ \ \ \ \ \ \ \ -\frac{rb^4\hbar^3}{16a^2\left(2\mu\right)^{3/2}}\left(r+\frac{\hbar^2}{2\mu a}\right)\left(\frac{8}{1575}b^{8} \alpha_0^8+\frac{1}{15}b^6\alpha_0^6+\frac{4}{9}b^4\alpha_0^4+b^2\alpha_0^2-\frac{F}{4ab^2}\right)^2\label{EQ20d}\\
E_{KH}^{(3)}&=&\int_{0}^{\infty}\mathcal{X}_{0}^2(r)\bigg[r^3 \left(\frac{8}{567}ab^{12}\alpha_0^8+\frac{128}{675}ab^{10}\alpha_0^6+\frac{4}{3}ab^8\alpha_0^4+\frac{32}{9}ab^6\alpha_0^2+\frac{4}{3}ab^4\right)\nonumber\\
&&\ \ \ \ \ \ \ \ \ \ \ \ \ \ \ \ \ \ \ \ \ \ \ \ \ \ \ \ \ \ \ \ \ \ \ \ \ \ \ \ \ \ -W_0^{(1)}(r)W_0^{(2)}(r)\bigg]dr\nonumber\\
&=&\frac{3\hbar^6}{128\mu^3a^3}\bigg[5\left(\frac{8}{567}ab^{12}\alpha_0^8+\frac{128}{675}ab^{10}\alpha_0^6+\frac{4}{3}ab^8\alpha_0^4+\frac{32}{9}ab^6\alpha_0^2+\frac{4}{3}ab^4\right)\nonumber\\
&&\ \ \ \ \ \ \ +\frac{9\hbar^4}{2048\mu^4a^4}\left(\frac{32}{1575}ab^{10} \alpha_0^8+\frac{4}{15}ab^8\alpha_0^6+\frac{16}{9}ab^6\alpha_0^4+4ab^4\alpha_0^2-F\right)^3\nonumber\\
&&\ \ \ \ \ \ \ \ \ \ \ \ -\frac{9\hbar^2}{64\mu a^2}\left(\frac{32}{1575}ab^{10} \alpha_0^8+\frac{4}{15}ab^8\alpha_0^6+\frac{16}{9}ab^6\alpha_0^4+4ab^4\alpha_0^2-F\right)\nonumber\\
&&\ \ \ \ \ \ \ \ \ \ \ \ \ \ \ \ \ \ \ \times\left(-\frac{8}{385}ab^{11}\alpha_0^8-\frac{112}{405}ab^9\alpha_0^6-\frac{40}{21}ab^7\alpha_0^4-\frac{24}{5}a b^5\alpha_0^2-\frac{4}{3}ab^3\right)\bigg]\label{EQ20e}
\end{eqnarray}
\end{subequations}
With Eqs. (\ref{EQ20a}-\ref{EQ20e}), we obtain the approximate energy eigenvalues and the wave function for an electron of helium atom under intense laser field encircled by an electric field as:
\begin{eqnarray}
E_{KH}\approx E_{KH}^{(0)}+\left(2ab-\frac{4}{405}ab^9\alpha_0^8-\frac{8}{63}ab^7\alpha_0^6-\frac{4}{5}ab^5\alpha_0^4-\frac{4}{3}ab^3\alpha_0^2\right)+E_{KH}^{(1)}+E_{KH}^{(2)}+E_{KH}^{(3)}+...,
\label{EQ21}
\end{eqnarray}
and
\begin{equation}
\Psi_0^{E_{_{KH}}}({r})\approx2\zeta^{3/2}r\exp{\left(-\zeta r\right)}\exp{\left(-\sqrt{\frac{2\mu}{\hbar^2}}\int_{}^r\left(W_{0}^{(1)}(\varrho)+W_{0}^{(2)}(\varrho)\right)d\varrho\right)},
\end{equation}
respectively. The behavior of energy eigenvalues  of an electron of helium atom interacting with electric field and exposed to linearly polarized intense laser field radiation as a function of various model parameters has been shown in Table 1 and Figure {\ref{fig2}}.

\begin{table}[h!]
{\footnotesize
\caption{Energy eigenvalues (in a.u.) of an electron of helium atom interacting with electric field and exposed to linearly polarized intense laser field radiation. we have taken $a=1$. } \vspace*{10pt}{
\begin{tabular}{p{0.6in}p{0.6in}p{0.6in}p{0.6in}p{0.6in}p{0.6in}p{0.6in}p{0.6in}p{0.6in}}\hline
{}&{}&{}&{}&{}&{}&{}&{}&{}\\[1.0ex]
$F$&0.0005&0.0010&0.0050&0.0100&0.0500&0.1000&0.2000&0.5000\\[3.5ex]\hline\\
$E_{KH}^{(b=0.01)}$&-7.9802&-7.9804&-7.9819&-7.9838&-7.9988&-8.0176&-8.0552&-8.1690\\[3.5ex]\hline\\
$b$&0.0004&0.0008&0.0040&0.0080&0.0400&0.0800&0.4000&0.8000\\[3.5ex]\hline\\
$E_{KH}^{(F=0.005)}$&-8.0011&-8.0003&-7.9939&-7.9859&-7.9219&-7.8420&-7.2139&-6.4659\\[3.5ex]\hline\hline\end{tabular}}}
\label{tab1}
\end{table}
\section{Numerical Result}
We delineate the effective potential and its approximate expression in figure \ref{fig1}. Specifically, Figures 1(a) and (b) show that intensifying the electric field strength will make the repulsiveness of the effective potential burgeon and Hence, the system becomes less attractive. It can also be seen that for a strong coupling strength and weak external electric field, the effective potential tends to continuum states. Furthermore, in Figure 1(c), by considering a weak external electric field, directed along $z$-axis, we found that the approximate expression to the model potential is only valid for low screening parameter.

We anticipate that the approximate expression may be influenced by variations in external electric field and coupling strength.  In order to elucidate this, we have plotted Figure 1(d). It can be seen that the approximate expression divaricates more from the model potential when $(a,F)=(5, 5)$ compares to when $(a,F)=(1, 0.01)$. This connotes that in order to obtain a better approximation to the potential model, we have to consider interaction with low electric field directed along the $z$-axis with $a=1$. However, it is worth mentioning that suppose the electric field is directed along $\theta=\pi$, a disparate conclusion might arise. Moreover, studying the variations in Figures 1(c) and (d), one can infer that the approximation is only valid for $br<<1$.

In Table $1$, we scrutinize the behavior of energy levels of an electron of helium atom as it interacts with electric field directed along $z$-axis and exposed to linearly polarized intense laser field radiation. It can be observed from this table that as the electric field becomes more intensify, the eigenvalues becomes more negative and the system becomes more repulsive. Moreover by considering a weak electric field and then varying the screening parameter, we found that the system becomes weakly bound. This corroborates the results of Figures 1(c) and 1(d) where we determine the validity condition for the approximate expression. In fact, one can predict that as screening parameter continues to increase, there will be a critical point where a transition from bound to continuum states takes place.

Furthermore, Figure 2(a) shows that the eigenvalue is inversely proportional to applied external electric field notwithstanding the choice of screening parameters (this can be seen from the slope of the graph). This figure reveals that for a very strong external electric field and an infinitesimal screening parameter, the system is strongly bound. Figure (2b) expounds this further. It can be seen that the eigenvalues increase monotonically with an increase of screening parameter for various intensities of external electric field. As the screening parameter burgeons, the system becomes weakly bond and highly repulsive. In fact this figure authenticate our projection in Table 1. At $b\geq3$, the system tends to continuum states.

In Figure 2(c), we examine the behavior of the eigenvalues as function of laser dressing parameter for a ginormous screening parameter and divers electric field intensities. It can be observed that as $\alpha_0$ increases, the $E_{KH}$ dwindles monotonically and becomes less repulsive. However, the scrutiny takes a different shape by considering a lilliputian screening parameter. Variation in the eigenvalues is indiscernible since the energy shift $\Delta E_{KH}=0$. In fact Figures 2(c) and (d) demonstrate the susceptibility of eigenvalues of an electron of helium atom to screening parameter. One can deduce that the eigenvalues will only response to variation in the frequency of the laser only if we consider ginormous screening parameter. But, this will invalidate our approximation, then we can only conclude that the variation in frequency of laser radiation has no effect on the eigenvalues of an electron of helium for a particular electric field intensity directed along $z$-axis.

\section{Concluding Remarks}
We scrutinize the behavior eigenvalues of an electron of Helium atom as it interacts with electric field directed along $\theta=0$ exposed to linearly polarized intense laser field radiation. In order to achieve this, one electron of the helium atom is frozen at its ionic ground state and the motion of the second electron in the ion core is treated via a more general case of screened Coulomb potential model.  Using the Kramers-Henneberger (KH) unitary transformation, which is a semiclassical counterpart of the Block-Nordsieck transformation in the quantized field formalism, the squared vector potential that appears in the equation of motion is eliminated and the resultant equation is expressed in KH frame. Within this frame, the resulting potential and the corresponding wave function have been expanded in Fourier series and using Ehlotzkys approximation, we obtain a laser-dressed potential to simulate intense laser field. By fitting the more general case of screened Coulomb potential model into the laser-dressed potential, and then expanding it in Taylor series up to $\mathcal{O}(r^4,\alpha_0^9)$, we obtain the solution (eigenvalues and wave function) of an electron of Helium atom under the influence of external electric field and high-intensity laser field, within the framework of perturbation theory formalism. It has been shown that the variation in frequency of laser radiation has no effect on the eigenvalues of an electron of helium for a particular electric field intensity but for a very strong external electric field and an infinitesimal screening parameter, the system is strongly bound. This work has potential application in the areas of atomic and molecular processes in external fields including interactions with strong fields and short pulses.  This work represents the continuation of our project ``atoms and molecule interacting with external fields in Laser-Plasma" which had commenced in refs. \cite{EF21,EF24,EF25}. We hope that, the current study will inspire furtherance in future by exploring the molecular system under intense laser field.

\section*{Acknowledgments}
We thank the referees for the positive enlightening comments and suggestions, which have greatly helped us in making improvements to this paper. This work was partially supported by 20160978-SIP-IPN, Mexico.

\end{document}